\documentclass[12pt]{iopart}

\usepackage{graphicx} % needed for figures %
\usepackage{dsfont}

\begin{document}

\title{Interaction of a propagating guided matter wave with a localized potential}

\newcommand{\LCARa}{Universit\'e de Toulouse; UPS; Laboratoire Collisions Agr\'egats R\'eactivit\'e (IRSAMC); F-31062 Toulouse, France}

\newcommand{\LCARb}{CNRS; LCAR UMR 5589 (IRSAMC); F-31062 Toulouse, France}

\newcommand{\LPTa}{Universit\'e de Toulouse; UPS; Laboratoire de Physique Th\'eorique (IRSAMC); F-31062 Toulouse, France}

\newcommand{\LPTb}{CNRS; LPT UMR 5152 (IRSAMC); F-31062 Toulouse, France}

\author{%
  G L Gattobigio$^{1,2}$, %
  A Couvert$^{1,2}$, %
  B Georgeot$^{3,4}$, %
  and %
  D Gu{\'e}ry-Odelin$^{1,2}$ %
} %
\address{$^1$\LCARa} %
\address{$^2$\LCARb} %
\address{$^3$\LPTa} %
\address{$^4$\LPTb} %

\ead{david.gueryodelin@gmail.com}
\begin{abstract}
We provide a theoretical framework to describe the interaction of a propagating guided matter wave with a localized potential in terms of quantum scattering in a confined environment. We analyze how this scattering correlates the  longitudinal and transverse degrees of freedom and work out analytically the output state under the Born approximation using a Gaussian localized potential. In this limit, it is possible to engineer the potential and achieve coherent control of the output channels. 
The robustness of this approximation is studied by comparing the stationary scattering theory to numerical simulations involving incident wave packets. It remains valid in a domain of weak localized potential that is achievable experimentally. We infer a possible method to determine the longitudinal coherence length of a guided atom laser. Then, we detail the non-perturbative regime of the interaction of the guided matter wave with the localized potential using a coupled channel approach. This approach is worked out explicitly with a square potential. It yields new non-perturbative effects such as the occurrence of confinement-induced resonances. The perspectives opened by this work for experiments are discussed.
\end{abstract}


\maketitle

The recent realization of guided atom lasers with a macroscopic
fraction of the atoms in the ground state of the transverse
confinement is a crucial step for the development of guided atom
optics \cite{GRG06, CJK08,GCJ09,DHJ09}. In such experiments the guided atom laser is produced by outcoupling the atoms from a trapped Bose-Einstein condensate (BEC) into an optical guide. 
In Ref.~\cite{GRG06}, the trap is magnetic and atoms are decoupled by radio-frequency spin flip. They propagate into an optical guide provided by a far-off resonance beam superimposed on the magnetic trap. The mean velocity of the beam is of the order of 10 mm.s$^{-1}$. Since the acceleration of atoms is due to the mean field of the condensate, the velocity dispersion is governed by the chemical potential. In Ref.~\cite{CJK08}, a BEC of rubidium-87 in a well-defined Zeeman sublevel $m_F$ is held in a far-off resonance crossed dipole trap. In this scheme, one of the two beams of the crossed dipole trap is nearly horizontal and is used as a guide. 
Relying either on the first-order ($m_F\neq 0$) or second-order Zeeman effect ($m_F= 0$) atoms are progressively outcoupled from the trap by increasing a magnetic field gradient applied along the direction of the guide (and thus decreasing the effective trap depth). Atoms are accelerated by the time dependent magnetic gradient yielding a large velocity dispersion depending on the instant at which atoms have been outcoupled. Alternatively, optical outcoupling was demonstrated \cite{GCJ09} by lowering the intensity of the vertical beam. The outcoupled atoms are slightly accelerated and acquire a mean velocity ranging from 10 to 20 mm.s$^{-1}$ after a 1 mm propagation. In this latter scheme, the strength of the potential experienced by the trapped atoms is progressively reduced which amounts to opening the trap adiabatically. At some point the adiabaticity breaks down and atoms are outcoupled into the guide. In all these experiments, there is a lack of information on the longitudinal degree of freedom and a need for methods allowing to measure the longitudinal velocity dispersion. 

In this article, we study the interaction of a guided atom laser
with an extra localized optical potential. The atom laser is modelised 
as a propagating matter wave without any atom-atom interaction
because of the diluteness of the experimentally realized guided
atom lasers \cite{GRG06,CJK08,GCJ09}. Such a study is of importance in the context of the exploration of the propagation of matter waves in complex structures. In addition, it offers a new possibility for measuring the longitudinal coherence length of the atom laser as explained below. 

The extra potential, also called ``defect'' in the following, can be generated by
a local modification of the transverse confinement such as a constriction \cite{JaS02,LCR03,LPS03}, or a
local curvature \cite{LeP01,BrE04}. Alternatively, one can use a far-off resonance light beam\footnote{A theoretical study with an on-resonance laser beam has also been carried out in Ref.~\cite{LMR06}.} superimposed on the guide to produce a large variety of shapes and strengths. This potential acts as an obstacle and enables one, for example, to revisit superfluidity for quantum degenerate guided beams \cite{Pav02}. 

The interaction between a guided matter wave and a localized potential belongs to the more general topic of elastic scattering in a multimode quantum waveguide, and has been studied in condensed matter for electronic matter waves
\cite{FGR90,Bag90,RRB93,Bee97,BLR99,KSJ99,CaM03,Gra04,LHR05}. The motivation there was to study the effect of
disorder on the electron transmission and, in particular, on
conductance quantization in narrow constrictions. 

Recent theoretical studies on atom-atom interaction where the true interacting potential is replaced by
the standard Huang-Fermi pseudo-potential have shown the occurrence of resonances in the two-body scattering length for atoms that are strongly transversally confined by a waveguide or more generally confined to quasi 1D geometry.  These confinement-induced resonances (CIR) for atoms have been predicted by Olshanii \cite{Ols98,BMO03}. A confinement-induced molecular bound state in a quasi-1D fermionic potassium-40 atom gas confined in a one-dimensional matter waveguide has been reported \cite{MSG05}.  However, the existence of the CIR in the scattering states has not yet been observed.

The regime studied here is beyond the limit of validity of the scattering length approach since (i) the potential range of the defect is of the order of the harmonic oscillator length associated with the transverse confinement (or even larger) and (ii) we do not restrict our approach to the low energy regime. Under these assumptions, one has to take into account the details of the interaction potential and it becomes possible to envision the tailoring of the defect potential to control the output channels after the interaction.

In Sec.~\ref{green}, we derive the theoretical framework
to describe the interaction between a dilute guided atom laser and a localized potential using the Green's function formalism in the presence of a transverse harmonic confinement \cite{Lupu98,LLH97,KSS05, PTM05}. This approach
yields an expression of the solution as a series expansion in the
powers of the strength of the localized potential. In Sec.~\ref{born}, we derive analytical results
for a weak Gaussian localized potential using the Born approximation. We also investigate the validity domain of the perturbative limit of the scattering results using numerical simulations. When the strength of the localized potential is increased, higher-order contributions become important. For this non-perturbative regime, we develop, in Sec.~\ref{matrix}, a matrix approach inspired by the theory of vibrational energy transfer in non-reactive collisions \cite{RaK68}. Using a model potential, we show how the coupling between external degrees of freedom that occurs in the localized potential region generates controlled entangled states. This process correlates the longitudinal wave vector to the transverse state, and yields new effects such as resurgences of quantum reflection.

 Without loss of generality, the theoretical description made in the following is done for a two-dimensional
problem; the matter wave propagates along the $x$ axis and the
transverse confinement, assumed to be harmonic with an angular frequency $\omega_\perp$, is provided along an
orthogonal direction, $y$. The corresponding Hamiltonian is therefore given by
\begin{equation}
H_0 = -\frac{\hbar^2}{2m}\left( \frac{\partial^2}{
\partial x^2}+\frac{\partial^2}{
\partial y^2} \right) + \frac{1}{2}m\omega_\perp^2 y^2. \label{eq1}
\end{equation}
The defect is taken into account as an extra
potential term $U(x,y)$, so that the total Hamiltonian reads $H=H_0 +U(x,y)$.

\section{Green's function formalism}\label{green}

To study the interaction of the guided atom
laser with the defect, we determine the scattering states by solving the stationary Schr\"odinger equation:
\begin{equation}
\left[H_0+U\right]|\varphi\rangle=E|\varphi\rangle.
\label{eqschrodinger}
\end{equation}
The formal solution of Eq.~(\ref{eqschrodinger}) is given by the Lippmann-Schwinger equation \cite{fetter}
\begin{equation}
|\varphi\rangle=|\varphi_0\rangle+
G^+U|\varphi\rangle,\,\,\mbox{where}\,\,G^+=\lim_{\epsilon \to
0}\frac{\mathds{1}}{E-H_0 + i\epsilon} \nonumber
\end{equation}
is the retarded propagator, $\mathds{1}$ is the identity matrix, and  $|\varphi_0\rangle$ a solution in
the absence of the defect $H_0|\varphi_0\rangle=E|\varphi_0\rangle$.

This formulation is well suited for a formal perturbative
expansion in powers of the localized potential $U$:
\begin{equation}
|\varphi\rangle=|\varphi_0\rangle+
G^+U|\varphi_0\rangle+G^+UG^+U|\varphi_0\rangle+ \ldots . \label{eq6}
\end{equation}

A natural basis of the Hilbert space for this scattering problem is provided by the
vectors $\{ |k,n \rangle = | k \rangle \otimes |n\rangle \}$, tensor products of the
longitudinal plane wave eigenvectors along the $x$ direction by
the eigenvectors of the harmonic potential associated with the
transverse degree of freedom. By
definition, one has $H_0 | k,n \rangle= E_{k,n} | k,n \rangle$,
with $E_{k,n}=\hbar^2k^2/2m+E_n$, where $E_n=(n+1/2)\hbar\omega_\perp$ is the
energy of the $n^{\rm th}$ level of the transverse harmonic
confinement. We introduce $\langle \vec{r} \,| k,n
\rangle= (2\pi)^{-1/2}e^{ikx}\psi_n(y)$, where $\psi_n(y)$
is the eigenfunction associated with the eigenenergy $E_n$.
The representation in position space of the retarded propagator is
by definition the Green's function, and its expansion on the $\{
|k,n \rangle \}$ basis reads \cite{MoF53}:
\begin{equation}
G^+(\vec{r},\vec{r}\,';E)  = \sum_{n=0}^\infty
\psi_n(y)\psi^*_n(y') g^+(x,x';E,n), \label{eq8}
\end{equation}
where $g^+(x,x';E,n)$ is the Green's function of an effective
one-dimensional scattering problem. Using the residue
theorem, one finds
\begin{equation}
g^+(x,x';E,n)=-\frac{mi}{\hbar^2} \frac{e^{ik_n(E)|x-x'|}}{k_n(E)},
\label{eq10}
\end{equation}
with $\hbar k_n(E)=\left[2m(E-E_n)\right]^{1/2}$. We obtain two kinds of modes: propagating ones, for which $k_n$
is real, and evanescent ones, for which $k_n$ is imaginary.

For an incident wave function of the
form $|k_0,0\rangle$, the interaction of the atom laser with the localized potential produces the contamination of the modes $|k_n,n\rangle$ in the
forward direction and $|-k_n,n\rangle$ in the backward direction. The asymptotic form of the wave function after interaction reads
\begin{eqnarray}
\varphi (x \to -\infty,y) & = & \langle
\vec{r} \,|k_0,0\rangle+\sum_{n \neq 0} r_n\langle \vec{r}\,|-k_n,n\rangle,
\label{eq14} \\ 
\varphi (x \to \infty,y) & = & \sum_{n} t_n\langle
\vec{r}\,|k_n,n\rangle, \label{eq13} 
\end{eqnarray}
where $t_n$ and $r_n$ are respectively the transmission and reflection amplitude coefficients in the $n^{\rm th}$ transverse mode. 

From Eqs.~(\ref{eq13}) and (\ref{eq14}) we find that the output wave resulting from the
interaction of the incident wave with the defect
is an entangled state, that is a linear superposition of correlated bipartite states involving both a transverse state of quantum number $n$ and a specific longitudinal state $\pm k_{n}$. By controlling the incident energy, one can choose the number of propagating modes, and thus the number of bipartite states that participate in the output state. 

We consider an incident wave packet built from a linear superposition of longitudinal wave vectors and transversally in the ground state of the confinement. After its interaction with the defect, the transmitted wave packet may undergo a kind of distillation in the course of its propagation. Indeed, as a result of energy conservation, for each incident wave vector $k$, the components with a non zero transverse quantum number $n$ have a reduced wave vector $k_n<k$:
\begin{equation}
\frac{\hbar^2k^2}{2m} + 0 = \frac{\hbar^2k_n^2}{2m} + n\hbar \omega_\perp.
\label{}
\end{equation}
 For a sufficient long propagation time, one
 therefore expects the packet to split into a sum of packets if the initial dispersion $\delta k$ in $k$ is small enough ($k\delta k \ll m\hbar \omega_\perp$).

\section{Born approximation}\label{born}

To determine the coefficients $r_n$ and $t_n$ for a weak localized potential, one applies the
Born approximation which consists in keeping only the first two
terms of the expansion of Eq.~(\ref{eq6}). Equation~(\ref{eq6}) combined with Eqs.~(\ref{eq8})
and (\ref{eq10}) yields the expression for the transmission and reflection
coefficients in the Born limit for $n>0$: 
\begin{equation}
\fl \qquad t_n  =  -\frac{2\pi mi}{\hbar^2|k_n|} \langle
k_n,n|U | k_0,0 \rangle \qquad {\rm and} \qquad
r_n  =    -\frac{2\pi mi}{\hbar^2|k_n|} \langle -k_n,n|U |
k_0,0 \rangle.
\label{eq16}
\end{equation}
These expressions are valid in the perturbative regime, $|t_n|\ll 1$ and $|r_n|\ll 1$.

\subsection{The Gaussian potential}

Let us apply the previous formalism to a model potential that explicitly couples the longitudinal and transverse degrees of freedom. We consider the case of a defect generated by a far-off resonance Gaussian beam:
\begin{equation}
U_g(x,y) = U_0 u_g(x/w_x)u_g(y/w_y) \ ,
\label{Gaussianpotential}
\end{equation}
where $u_g(x)={\rm e}^{-2x^2}$ and $w_x$ (resp.~$w_y$) is the waist along the $x$ (resp.~$y$) direction. By combining Eqs.~(\ref{eq16}) and (\ref{Gaussianpotential}), one finds: 
\begin{equation}
\fl t_{2p}(k_{2p})  =  -i  \frac{U_{0}}{\hbar\omega_{\perp}} \sqrt{\frac{\pi}{2}} 
\frac{%
  {\rm e}^{%
    -( k_{2p}-k_{0})^{2}
    \frac{w_x^{2}}{8}
  }}%
  {\left|k_{2p}\right|a_{\rm ho}^2}
\, w_x
\, g_{2p}(\eta),\,\,\,\, {\rm and} \,\,\,\,
r_{2p}(k_{2p}) = t_{2p}(-k_{2p}) \ ,
\label{tn-et-rn-gaussien}
\end{equation}
where $p$ is an integer, $\eta=w_y/2a_{\rm ho}$, $a_{\rm ho}=(\hbar/m\omega_\perp)^{1/2}$ is the harmonic oscillator length, and $g_{2p}(\eta)$ accounts for the matrix element of the potential $U(x,y)$ between the transverse oscillator states (see Appendix A):
\begin{equation}
g_{2p}(\eta) =  \langle 2p | u_g(y) | 0 \rangle= (-1)^p \frac{\sqrt{2(2p)!}}{2^{p}\,p!}
\, \eta
\, \left( \frac{1}{1+2\eta^2} \right)^{p+\frac{1}{2}} \ .
\label{eqg1peta}
\end{equation}
The parity of the potential with respect to the $y$ variable cancels out the contributions of the odd terms.

\subsection{Validity of the perturbative approach}

The predictions of the Born approximation can be compared with the results
of direct simulations of the scattering of wavefunctions by such Gaussian potentials. 
Indeed, an incident ideal guided matter wave can be modeled by a wave packet or a statistical mixture of wave packets \cite{CDM00}.
We thus use Gaussian packets as initial state for the numerical simulations.
In order to test the robustness of the results of the Born approximation that involves an incoming plane wave, we have chosen the wave packet $\pi(x)$ with a half longitudinal size, $L_p$, defined as $\pi(x)=|\psi(x)|^2=(2\pi)^{-1/2}L_p^{-1}\exp(-x^2/2L_p^2)$. 
The time evolution of the wave packet is performed using the split-operator technique, for which the evolution
operator is approximated by the product of the potential and kinetic term
for a succession of small time steps.
The results depend on several parameters, in
particular the strength of the potential
(proportional to the power, $P$, of the laser used to generate the defect) 
and its width $w=w_x=w_y$. 

 The wave packet is launched towards an attractive defect, then interacts with it and the projections on the different transverse modes are eventually computed.
Initially, the wave packet is assumed to be entirely in the transverse ground state $n=0$.
For the simulations, we consider a matter wave of rubidium-87 atoms propagating, with a mean velocity of $10$ mm.s$^{-1}$, in a guide provided by a far-off resonance dipole beam, $\lambda=1070$ nm, with a waist of $45$ $\mu$m. The defect
corresponds to a Gaussian potential such as (\ref{Gaussianpotential})
with $U_0/k_B=9.75 \times 10^{-2}  P/w^2$ K where $P$ is the power in Watts and $w$ the
waist in $\mu$m.
These values are consistent with recent experiments \cite{GRG06, CJK08,GCJ09}.  

\begin{figure}[t]
  \centering
\includegraphics[scale=0.85,angle=-0]{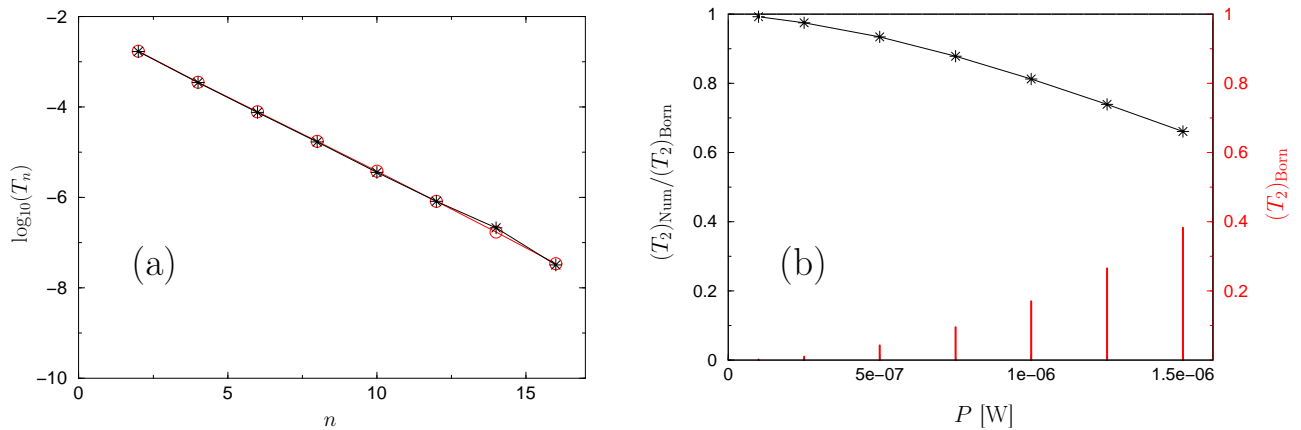}
  \caption{(a) Logarithm (decimal) of the probability of transmission $T_n=k_n|t_n|^2/k_0$ as a function of the transverse mode number $n$ of the guide (even states only): Comparison between the numerical simulations (black stars) and the Born approximation results (red open circles). The initial wave packet of longitudinal width $L_p=22$ $\mu$m and in the transverse mode
$n=0$ of the guide potential moves towards an attractive Gaussian defect (waist $w=1$ $\mu$m and power $P=10^{-7}$ W) at a mean velocity of $10$ mm.s$^{-1}$. The guide generated by a dipole beam of waist $45$ $\mu$m has an angular frequency $\omega_\perp= 2\pi \times 203$ Hz. (b) Ratio between the probabilities of transmission in the transverse state $n=2$
 obtained by the numerical simulations $(T_2)_{\rm Num}$ and deduced from the Born approximation $(T_2)_{\rm Born}$ (black stars), as a function of the power of the defect $P$ in W (other parameters are as in (a)).
Red bars denote the value of $T_2$ given by the Born approximation.}
  \label{Borntest1}
  \end{figure}

For small $P$ and $w$, the projection of 
the scattered wavefunction on the different transverse modes is in very good agreement 
with the predictions of the Born approximation.  Figure \ref{Borntest1}(a) shows 
that for $P=10^{-7}$ W and $w=1$ $\mu$m the projections on the first 16 transverse modes
are well reproduced.  Such widths of the defect, although small,
 are within reach of current experimental possibilities.  

We have investigated the validity of the Born approximation both as a
function of $P$ and $w$. As concerns the power dependence, the role of the  higher-order terms in the expansion (\ref{eq6}) is clear since the localized potential is simply proportional to the power. For instance, we find that, for small widths of the defect such as $w=1$ $\mu$m and $w=4$ $\mu$m, the results of Born approximation remain approximately correct for the first transverse mode up to values 
of $P$ where a significant part of the atoms ($\approx 40 \%$) is transferred
from $n=0$ to $n=2$ (see Fig.\ref{Borntest1}(b)).

 The dependence of the higher-order terms of the expansion (\ref{eq6}) on the width of the localized potential is less simple.
We first note that in view of the formula (\ref{tn-et-rn-gaussien}),
the Born prediction for a given value of $P$
makes sense only in a certain range of values of $w$: for small $w$ it
predicts projections too large to be valid (eventually larger than one), 
while for large $w$
the predictions are so small that they are practically useless.
The intermediate regime corresponds to a waist of the order of $12$ $\mu$m with $P=10^{-3}$ W, 
and of $20$ $\mu$m with $P=10^{-1}$ W.
Generally,  our numerical simulations show that for larger values of the width, the agreement of Born predictions with the numerics is less good than for smaller width even when a comparable fraction of atoms is transferred (data not shown). This indicates that the higher-order terms in the expansion (\ref{eq6}) grow faster with the size $w$ than with the power $P$ of the Gaussian potential.

\subsection{Coherent control in the perturbative limit}

When the waist of the Gaussian potential is much larger than the oscillator length, $\eta\gg1$, the excited levels of the transverse confinement are nearly not populated. In order to control the population of the output scattering states, the defect size should thus be of the same order as the oscillator length ($1$~$\mu$m for typical experimental conditions \cite{GRG06,CJK08, GCJ09}). 

By engineering  the localized potential, it is possible to control the population in the different output channels.  For instance, the following family of potentials
\begin{equation}
U_m(x,y) = U_0 u_g(x/w_x)u_g(y/w_y)
\,[ H_{m}(y/a_{\rm ho})+C] \ ,
\label{pot-hermite}
\end{equation}
where $H_{m}$ is the Hermite polynomial of order $m$ and $C$ a constant, permits one to populate only the $m^{\rm th}$ mode of the transverse guide in the large waist limit ($\eta\gg1$). Indeed, the $g_p$ coefficients are given for $p\neq 0$ by
\begin{equation}
\fl g_p(\eta)  = \langle p | u_g(y/w_y) [ H_{m}(y/a_{\rm ho})+C] | 0 \rangle \propto \langle p | u_g(y/w_y)  | m \rangle \simeq  \langle p  | m \rangle= \delta_{mp} \ ,
\end{equation}
where $\delta_{mp}$ is the Kronecker delta function. A direct consequence is the possibility of creating propagating states in a linear superposition of transverse states by using an interacting potential of the form (\ref{pot-hermite}) where the Hermite polynomial $H_{m}$ is replaced by a sum of Hermite polynomials.

\section{Coherence length of a guided atom laser}

The coherence properties of a guided continuous atom laser have been
investigated theoretically in Ref.~\cite{CDM00}. This study assumed the propagating beam to
be transversally confined (with two transverse dimensions $x$ and $y$) and at
thermal equilibrium in the frame moving at its mean velocity. However, it is
not clear to which extent this thermodynamical model can be applied to the
guided atom lasers generated so far by outcoupling atoms from BECs. Indeed, the thermalisation process involves collisions
with atoms outside the condensate, and can be very long in our case.

\begin{figure}[t]
  \centering
\includegraphics[scale=0.5,angle=-0]{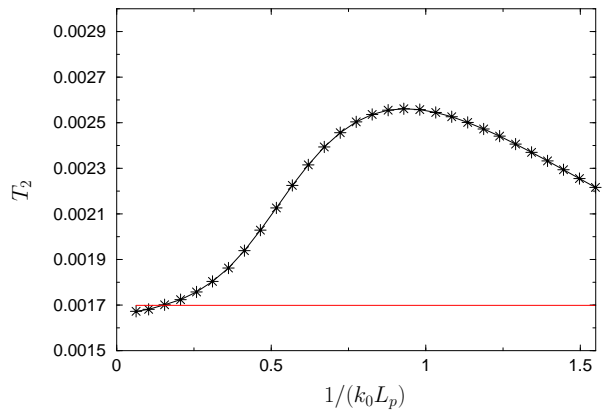}
  \caption{Transmissions $T_2$ to the transverse state $n=2$
 from the numerical simulations (black stars), as a function of the dimensionless quantity $1/(k_0L_p)$, for  a width of the transverse Gaussian potential of $w=1$ $\mu$m and power $P=10^{-7}$ W.
Except for $L_p$, the initial wave packet parameters are the same as in
Fig.~\ref{Borntest1} with a mean velocity of 10 mm.s$^{-1}$. The red horizontal line corresponds to the prediction of the Born approximation for an incoming plane wave of momentum $\hbar k_0$. Scattering is analyzed after the same evolution time for each $L_p$.}
  \label{Borntest3}
  \end{figure} 

In the following, we simply assume that the condensate is well-described by a Gaussian
wave packet moving in the waveguide. The scattering of a wave packet can be described by the
scattering of a coherent superposition of plane waves as studied in Sec. 1. When the dispersion in
momentum $\delta k$ of the wave packet is small compared to its mean wave
vector $k_0$, the scattering turns out to be well described by the Born approximation result for a plane wave of wave vector $k_0$ \footnote{In Sec.~2, we have used this regime of parameters.}. The size of the wave packet $L_p= \delta k^{-1}/2$ is considered as the coherence length.

In subsection 2.2 we considered the validity of the Born approximation
as a function of the power $P$ and waist $w$. Let us now
discuss its domain of validity as a function of $L_p$.
We concentrate on the first excited state (most populated). When the kinetic energy in this excited state tends to zero the Born approximation breaks down because of the $1/k_{2}$ dependence of the reflection and transmission coefficients (see Eqs.~(\ref{tn-et-rn-gaussien})). This affects significantly the wave packet scattering. Indeed, in Fourier space, depending on the dispersion $\delta k$,
some components of the wave packet in momentum may reach this regime.
In this case, one can distinguish three different zones of wave vectors $k$: those sufficiently large so that the Born approximation for $k_0$ remains valid, those for which the incident energy is not sufficient to populate any excited state, and an intermediate zone where the Born approximation first increases the scattering to the excited state then breaks down because of the divergence in $1/k_{2}$. 
When $\delta k$ increases, one thus should expect an enhancement of the occupation of the excited state when the intermediate zone becomes 
significantly populated.  A further increase of $\delta k$ should eventually at some point decrease the population of the excited state as the momenta when no scattering occurs have more and more weight in the wave packet.

Figure~\ref{Borntest3} shows the results of the simulation of wave packets
of different sizes compared to the Born approximation.  One
clearly sees an increase in the population of the excited state by more than 50 \% starting from
 a longitudinal length $L_p\approx 0.35$ $\mu$m.  This increase is followed by
a decrease for smaller values of $L_p$, as expected from the above qualitative discussion. This
phenomenon allows to estimate the coherence length of the wave packet by tuning the
parameters of the system until one reaches the breakdown of the Born
approximation. In this non-stationary regime, the precise maximum value obtained
for $T_2$ depends on the time at which the observation is made, since a part
of the wave packet has very small velocity components. 
The accuracy of determination of the coherence length of the atom laser through its interaction with a defect depends on the value of 
$k_0$.  Indeed, the breakdown of the Born approximation takes place
when values of $k$ corresponding to very small $k_{2}$ are present.
The threshold is thus given by $\hbar^2 (k_0-\delta k)^2/2m \approx E_2$
or $k_0-\delta k \approx \sqrt{5}/a_{\rm ho}$.
In the regime corresponding to the experiments 
\cite{CJK08,GCJ09}, one has 
$E_2 \ll \hbar^2 k_0^2/2m$, so that to have momentum
components with a small $k_2$ one needs $k_0 \approx \delta k$.  Thus
the range of $L_p$ values that can be probed depends on the range of initial
velocities (which depends in turn on the outcoupling mechanism).

In summary, there is a regime of parameters in $k$ for which the scattering strongly depends on the longitudinal size $L_p$ of the wave packet. The breakdown of the Born approximation in this regime may enable to probe the coherence length ($\propto (\delta k)^{-1}$) of the atom laser from measurements of the population of the transverse modes.

Other techniques, developed for quasi-condensates in very elongated
geometry such as Bragg
spectroscopy \cite{RGT03} or ballistic expansion \cite{HDR01} could be  envisioned to determine the coherence length. The temporal coherence of an atom laser beam has been
experimentally investigated in \cite{kohl} using the interference contrast of the standing wave pattern that results from the reflection on a potential barrier. However, this latter technique essentially provides an upper limit on the coherence time.
The method exposed here shares some similarities with the proposal in
\cite{carusotto} in which the matter wave scatters on a finite optical 
lattice. However, this proposal is derived assuming a purely one-dimensional system. In contrast, we take into account and even make full use of the excitation of the transverse modes after the interaction with the localized potential.

\section{Non-perturbative treatment}\label{matrix}

The results presented so far are valid only within the Born approximation. 
However, this perturbative approach is valid only in a limited regime of parameters (see sections 2 and 3), and cannot account for resonance
phenomena such as confinement induced resonances and Fabry-Perot resonances.
In this section, to address this non-perturbative regime, we detail a matrix approach inspired by the theory of vibrational energy transfer between simple molecules in non-reactive collisions \cite{RaK68} or similarly by inelastic quantum
scattering theory \cite{child}. This method is non-perturbative with respect to the potential strength $U_0$, and therefore yields the non-perturbative response of the atom laser propagation.

\subsection{Matrix formalism}

This approach consists in (i) expanding the
scattering wave solution in terms of the transverse harmonic oscillator wave
functions as $\psi(x,y)=\sum_n \phi_n(x)\psi_n(y)$, and (ii)
taking the scalar product of the Schr\"odinger equation in
the presence of the localized potential with a given $m$ state of the
transverse harmonic oscillator. In this way, we find that the
longitudinal functions $\phi_n(x)$ obey a set of coupled second-order
linear equations:
\begin{equation}
\left[ \frac{d^2}{d x^2} + k_m^2 - \tilde{U}_{m,m}(x)\right]
\phi_m(x)=\sum_{n\neq m} \tilde{U}_{m,n}(x)\phi_n(x)
\label{eq20}
\end{equation}
with $\tilde{U}_{m,n}(x) = 2m \langle m |U(x,y)|n\rangle/\hbar^2$.
The terms involving only the $m$ index describe the ``elastic''
component of the scattering, while the term which involves
$\tilde{U}_{m,n}$ (with $m\neq n$) is responsible for the coupling
channel dynamics between the different transverse oscillator
states. The coupling between these 1D problems due to the right hand side terms of Eq.~(\ref{eq20})
gives rise to new resonances, as discussed in Sec.~\ref{qrr} on a particular example, in close analogy with Feshbach resonances \cite{KGJ06}.

The set of Eqs.(\ref{eq20}) must be solved subject to the
asymptotic conditions  (equivalent to Eqs.~(\ref{eq13}) and
(\ref{eq14})): 
\begin{eqnarray}
\phi_n (x \to -\infty) & = & \langle x|k_0\rangle \delta_{n,0}+
r_n\langle
x|-k_n\rangle, \label{eq21a}\\
\phi_n (x \to \infty) & = & t_n \langle x|k_n\rangle. \label{eq21b}
\end{eqnarray}
Through conservation of the probability current, the coefficients $r_n$ and $t_n$ are related by 
\begin{equation}
k_0 = \sum_{n=0}^{N_p}k_n \left(|r_n|^2+|t_n|^2\right), \label{eq22}
\end{equation}
where ${N_p}$ denotes the number of propagating modes. Such a relation
is particularly useful for checking the validity of numerical
solutions, and permits one to define the level of contamination of
a transverse ``vibrational'' state, $n$, both for the reflection
$R_n=k_n|r_n|^2/k_0$ and the transmission
$T_n=k_n|t_n|^2/k_0$ with the normalization rule $
\sum_{n=1}^{N_p}\left(R_n+T_n\right)=1$.

\begin{figure}[t]
  \centering
\includegraphics[scale=0.8]{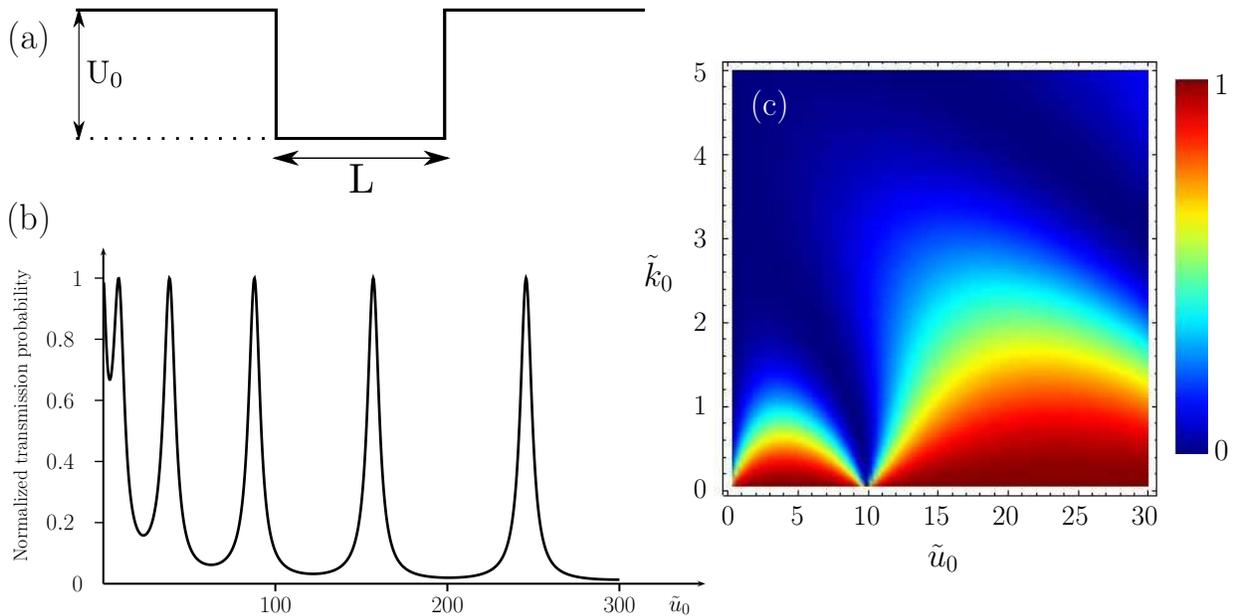}
  \caption{(a) 1D square well potential.
  (b) Probability of transmission $T$ for a 1D square potential for $\tilde{k}_0=k_0L=1$ as a function of the normalized depth
$\tilde{u}_0=2mL^2U_0/\hbar^2$. The transmission peak resonances are the analogous of the Fabry-Perot effect in optics. (c) Reflection probability for a one-dimensional square well of length $L$ and depth $U_0$ as a function of the incident wave vector $\tilde{k}_0=k_0L$ normalized to $L$ in ordinate and the potential depth $\tilde{u}_0$. Total reflection occurs when the incident energy is low compared to the potential depth.}
  \label{fig3}
  \end{figure}

\subsection{Application to the square well potential}

In order to analyze the scattering in the presence of open channels, we compare in the following the results obtained
 from the matrix model involving a 2D square well potential with its well-known counterpart in 1D.

\subsubsection{Reminder of the 1D scattering problem}
\ \\
Let us consider a one-dimensional square potential well of length $L$ and depth $U_0$ (see Fig.~\ref{fig3}(a)). The total reflection probability $R$ for an incident wave vector $k_0$ reads \cite{messiah}
\begin{equation}
R(\tilde{k}_0,\tilde{u}_0) = \frac{(\tilde{k}_0^2-\tilde{q}^2)^2\sin^2\hat{q}}{(\tilde{k}_0^2+ \tilde{q}^2)^2\sin^2 \tilde{q}+4 \tilde{k}_0^2 \tilde{q}^2\cos^2 \tilde{q}},
\label{QR1}
\end{equation}
where we have introduced the dimensionless parameters $\tilde{k}_0=k_0L$, $\tilde{u}_0=2mL^2U_0/\hbar^2$, and $\tilde{q}=(\tilde{k}_0^2+\tilde{u}_0)^{1/2}$.
The transmission $T=1-R$ plotted as a function of $\tilde{u}_0$ for $\tilde{k}_0=1$ exhibits Fabry-Perot resonances for which the incident wave is perfectly transmitted without any reflection (see Fig.~\ref{fig3}(b)). 
A total quantum reflection due to the sharp edges of the square well potential is obtained when the incident energy is low compared to the potential depth, $\tilde{k}_0^2\ll \tilde{u}_0$. This result is illustrated in Fig.~\ref{fig3}(c) where the total reflection probability $R$ is plotted as a function of $\tilde{k}_0$ and $\tilde{u}_0$. 

\subsubsection{Results for the 2D scattering problem}\label{qrr}
\ \\

\begin{figure}[t]
  \centering
  \includegraphics[scale=0.8]{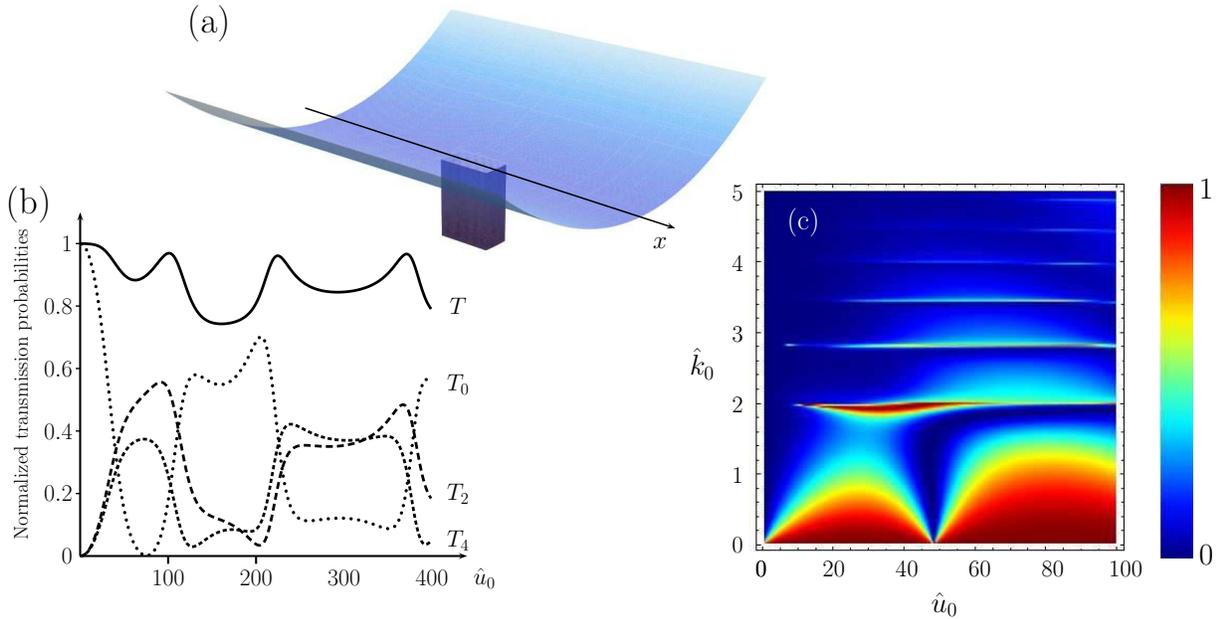}
  \caption{
  (a) 2D square potential superimposed on the transverse harmonic confinement.
  (b) Total probability of transmission $T=\sum_n T_n$ (solid line), and the probabilities of transmission $T_n$ in the transverse state $n$ (dotted and dashed lines), as a function of the normalized depth $\hat{u}_0=2ma_{\rm ho}^2U_0/\hbar^2$ with the incoming longitudinal wave vector $k_0=3.3/a_{\rm ho}$ (see Eq.~(\ref{eq17}) with $L_x=a_{\rm ho}$ and $L_y=0.1\,a_{\rm ho}$). In this example, there are $N_p=3$ even propagating modes.
  (c) Total reflection probability $R=\sum_n R_n$ for a 2D scattering problem with
 a harmonic confinement in the transverse and a localized potential of the form (\ref{eq17}) as a function of the incident
 wave vector, $\hat{k}_0=k_0a_{\rm ho}$, normalized to the oscillator length $a_{\rm ho}=(\hbar/m\omega_\perp)^{1/2}$ on the vertical axis and the potential depth $\hat{u}_0$. Resonances in the reflection probabilities are observed for discrete values of the incident wave vector: $k^{(n)}_0=(2n/a_{\rm ho})^{1/2}$, they correspond to resurgences of quantum reflection due to the coupling between transverse and longitudinal degrees of freedom.}
  \label{fig4}
  \end{figure}

Let us consider now a 2D square well potential (see Fig.~\ref{fig4}(a)) 
\begin{equation}
U(x,y)=-U_0u_b(x;L_x)u_b(y+L_y;2L_y), \label{eq17}
\end{equation}
where $u_b$ is the square function $u_b(x;L_x)=\Theta(x)\Theta(L_x-x)$ with $\Theta(x)$ the Heaviside
function superimposed on the transverse harmonic confining potential of the guide. The localized potential is chosen to be
 transversally centered. This model potential enables simple calculations and captures important features associated with the coupling between the
longitudinal and transverse degrees of freedom. In the regions $x\le 0$ and $x
\ge L_x$, the wavefunction $\phi_n$ is given by Eqs.~(\ref{eq21a}) and (\ref{eq21b}) respectively. In the region
of the localized potential, one has to solve the equation:
\begin{equation}
\frac{{\rm d}^2\vec{f}}{{\rm d}x^2}+\mathsf{M}  \vec{f}=0, \label{eq23}
\end{equation}
where $\vec{f}$ is the vector of coordinate
$(\phi_0,\phi_1,\ldots,\phi_n,\ldots)$ and $\mathsf{M}$ the matrix of elements
$\mathsf{M}_{n,n'}=k_n^2\delta_{n,n'}-\tilde{U}_{n,n'}$. The method used to solve this linear set of equations with the potential (\ref{eq17}) is detailed in Appendix B.

As an example, we have plotted in Fig.~\ref{fig4}(b) the total transmission probability $T$ along
with the transmission probabilities, $T_n$, in each propagating mode
($n=0,2$ and $4$) as a function of the normalized depth $\hat{u}_0=2ma_{\rm ho}^2U_0/\hbar^2$. In this example, the incident wave vector is $k_0=3.3/a_{\rm ho}$ (therefore there are only $N_p=3$ propagating modes) and the potential is chosen of the form Eq.~(\ref{eq17}) with $L_x=a_{\rm ho}$ and $L_y=0.1a_{\rm ho}$.

The interaction of the matter wave with the localized potential
leads to nearly total transparency for some discrete
specific values of the depth of the well ($\hat{u}_0\simeq-101,\,-225,\,-371$ in Fig.~\ref{fig4}(b)) as in the 1D case. 
In addition, new resonances occur with, for
instance, a maximum probability of transmission $T_2$ in the
transverse oscillator state $n=2$. An example is provided in Fig.~\ref{fig4}(b) for the
transmission probabilities at
$\hat{u}_0\simeq -92$, where less than 5\% of
the wave is reflected and for which $T_0 \simeq 0.07$,
$T_2 \simeq 0.55$ and  $T_4 \simeq 0.33$. At the exit of the
localized potential region, the state of the guided atom laser is
a linear superposition of the propagating eigenvectors
$|k_n,n\rangle$. 

The coupling between the longitudinal and transverse degrees of freedom along with the two characteristic lengths of the square well potential yield a quite complicated structure for the transmission probabilities.
Narrow resonances with nearly no transmission may occur at very large depths. However, they depend on the details of the potential.
To get a more complete understanding of the parameters for which total reflection may occur, we have plotted in Fig.~\ref{fig4}(c) the total reflection probability $R=\sum_nR_n$ as a function of the normalized incident wave vector, $\hat{k}_0=k_0a_{\rm ho}$, and the potential depth, $\tilde{u}_0=2ma_{\rm ho}^2U_0/\hbar^2$. In Fig.~\ref{fig4}(c), one observes the appearance of new reflection resonances for a set of discrete values of the incident wave vector, $k^{(n)}_0=(2n/a_{\rm ho})^{1/2}$, that correspond to the opening of a new transverse channel accessible by the incident energy. The full understanding of the detailed structure of the transmission probabilities would require further investigation.

One understands the physical origin of these resonances as follows: when the incident wave enters the localized potential zone, there is a strong
coupling between transverse and longitudinal degrees of freedom. This coupling favors a virtual excitation of the transverse excited modes so that the longitudinal energy is, again, in a range for which total reflection would occur in 1D. This effect is reminiscent of the confinement-induced resonances
\cite{Ols98,BMO03,PTM05,SMS08}.

\section{Discussion and conclusion}

The calculations described in this article are motivated by the recent realization of guided atom lasers in the ground state of the transverse confinement \cite{GRG06, CJK08,GCJ09, DHJ09}. The interaction of the matter wave with a localized potential where open and closed channels are accessible through the scattering of the matter wave on the potential is in direct analogy with non-reactive chemical reaction and guided electronic wave propagation in mesoscopic physics.

The possibility of shaping the localized potential using established cold atom techniques opens new perspectives:
\begin{itemize}
\item It allows for the generation and the manipulation of matter wave entangled states that correlate longitudinal and external degrees of freedom. This description in terms of entangled states is well-suited for the analysis and interpretation in the presence of a succession of interacting zones. Such states can be envisioned to encode or stock information in the external degrees of freedom, or in both internal and external degrees of freedom. This latter prospect is reminiscent of the use of laser light for quantum computation with cold trapped ions \cite{CiZ95}.
\item It provides a new method to determine the coherence length of a guided atom laser without making the assumption of the thermodynamical equilibrium. 
\item It constitutes a new device to investigate coherent control of the population of the discrete energy levels through their interaction with the localized potential\footnote{The control of the
transmission output channels implies the shaping of the localized
potential on the typical scale of the harmonic oscillator length
corresponding to the transverse confinement. Another advantage of
a very local action of the extra potential lies in the fact that
the reflection or transmission effects become less sensitive to
the de Broglie wavelength dispersion when the size of the
potential decreases, and are therefore more robust with respect to the
longitudinal monochromaticity of the incoming guided atom laser.}.
\item It provides a test bed for inverse scattering problem in confined environments \cite{invscatpro,GeR01}. It allows, in principle, to engineer potential with different shapes but with the same output after the interaction. It therefore enables the investigation of the functional relationship between the potential details and the scattering properties.
\end{itemize}

A natural extension of this work also lies in the inclusion of atom-atom interactions. Many new effects are expected. For
instance, the propagation of an interacting beam through a constriction
\cite{LCR03} (local increase in the strength of the transverse
confinement) has no stationary solution if the compression is
sufficiently high, but solitonic like solution \cite{LPS03}.
This effect arises from the nonlinearity of the mean field term
that describes the interactions. Other effects are connected to
the quantum turbulent regime downstream the obstacle realized by
the localized potential \cite{Berloff,HSR09}. Nonlinear atomic optical
effects are expected to arise from atom-atom collisional
interactions in a single-mode atomic Fabry-Perot cavity driven by
a coherent cw atom laser beam \cite{Car01}. The role of
interaction within the mean field description on quantum
scattering problems is a fundamental issue related to the physics of nonlinear Schr\"odinger equation that can be addressed with guided atom lasers \cite{RaK08}.

\ack
It is a pleasure to thank Yvan Castin and Gonzalo Muga for useful discussions.
We thank Renaud Mathevet et Thierry Lahaye for useful comments. We are grateful to Duncan England for careful reading of the manuscript. We acknowledge financial support from the Agence Nationale de la Recherche (ANR-09-BLAN-0134-01) and the R\'{e}gion Midi-Pyr\'{e}n\'{e}es. We thank CalMip for the use of their supercomputer facilities.

\section*{Appendix A}

In the Born approximation, the characterization of the interaction of a matter wave with a localized potential 
through the reflection and transmission amplitude coefficients requires 
the knowledge of the quantity $ \langle k_{n}, n| \,U\,|k_{0},0 \rangle$ (see Eq.~\ref{eq16}). 
In this Appendix, we detail the calculation of this matrix element in the case of the Gaussian localized potential (\ref{Gaussianpotential}):
\begin{equation}
\fl\langle k_{n}, n| \,U\,|k_{0},0 \rangle = \frac{U_{0}}{2\,\pi}\int_{-\infty}^{+\infty}e^{-i(k_{n}-k_{0})\,x}\,
e^{-2x^2/w_x^2} {\rm d}x 
\underbrace{ \int_{-\infty}^{+\infty}\psi_{n}^{*}(y)\,\psi_{0}(y)e^{-2y^2/w_y^2} {\rm d}y
 }_{\displaystyle{g_{n}}  }.\label{eq:A1}
\end{equation}
This expression contains two factors, the Fourier transform of the longitudinal
Gaussian potential $u_{g}(x/w_x)$ which can be easily calculated, and an integral $g_{n}$ which involves the
eigenfunctions of the harmonic oscillator, defined by:
\begin{equation}
        \psi_{n}(y)=\frac{1}{\sqrt{a_{\rm ho}\,
        n!\,2^{n}\,\sqrt{\pi}}}H_{n}\left(\frac{y}{a_{\rm ho}}\right)\,e^{-y^2/2a_{\rm ho}^2}        \label{eq:A2}
    \end{equation}
where $H_{n}(u)$ are the Hermite polynomials of order $n$.
The integral $g_{n}$ can therefore be rewritten as
\begin{equation}
    g_{n}\,=\,\int_{-\infty}^{+\infty} \frac{ {\rm d}u}{\sqrt{2^n\,n!\,\pi}}\, 
    H_{n}(u)\, e^{-\,u^2\,\alpha^{2} }\,,
    \label{eq:A3}
\end{equation}
where we have introduced the two dimensionless parameters $\eta = w_{y}/(2a_{\rm ho})$ and
$\displaystyle{\alpha^{2}=1+1/(2\,\eta^2)}$.

Let us calculate explicitly the generating function, $f(z)$, associated with the integrals $g_{n}$:   
\begin{eqnarray} 
\fl f(z) & = & \sum_{n=0}^{+\infty}g_{n}\frac{z^{n}}{n!}
=\frac{1}{\sqrt{2^n\,n!\,\pi}}\,\int_{-\infty}^{+\infty}\left(\sum_{n=0}^{+\infty}H_{n}(u)
    \frac{z^{n}}{n!}\right)e^{-\alpha^{2}u^{2}} {\rm d}u \nonumber\\ 
\fl    & = &  \frac{1}{\sqrt{2^n\,n!\,\pi}}\,\int_{-\infty}^{+\infty}e^{2uz-z^2}e^{-\alpha^{2}u^{2}}
    {\rm d}u 
    = \frac{1}{\sqrt{2^n\,n!}}\,
    \frac{1}{\alpha}
    \sum_{n=0}^{+\infty}
    \frac{z^{2n}}{n!}\left(\frac{1}{\alpha^{2}}-1\right)^{n}
 \label{eq:A4}\,. 
\end{eqnarray}   
Identifying the different terms of this last Taylor expansion with those of the definition of 
the generating function we deduce the expression for the integral, $g_{n}$:
\begin{equation}
g_{2p}(\alpha) =  \frac{1}{\sqrt{\pi(2p)!2^{2p}}}\frac{\sqrt{\pi}}{\alpha}\frac{(2p)!}{p!}
 \left(\frac{1}{\alpha^{2}}-1\right)^{p}, \,\, { \rm and}\,\,     g_{2p+1}(\alpha)  =  0,
\label{eq:A9}
\end{equation}
where $p$ is an integer. This proves Eq.~(\ref{eqg1peta}). Finally, we find: 
\begin{equation}
 \fl   \langle k_{n}, n| \,U_{0}\,u_{g}(x/w_x)\,u_{g}(y/w_y)\,|k_{0},0 \rangle =
    \frac{U_{0}}{2\pi}\sqrt{\frac{\pi}{2}}\,w_{x}\,
    e^{(k_{2p}-k_{0})^2\,\frac{w_{x}^2}{8}}\,g_{2p}(\eta).
\end{equation}

\section*{Appendix B}

To find the solution of Eq.~(\ref{eq23}), one needs to diagonalize the matrix $\mathsf{M}$. Let us introduce the matrix $\mathsf{P}$ such that $\mathsf{M}=\mathsf{P}  \mathsf{D}  \mathsf{P}^{-1}$ and
the vector $\vec{\ell} = \mathsf{P}^{-1}  \vec{f}$, Eq.~(\ref{eq23}) can be
rewritten in the diagonal form $\ddot{\vec{\ell}}+D\vec{\ell}=0$ whose solution reads
\begin{equation}
\vec{f}(x)=\mathsf{P}  e^{i\mathsf{D}^{1/2}x}  \vec{a}+\mathsf{P}
e^{-i\mathsf{D}^{1/2}x}  \vec{b}, \label{eq23b}
\end{equation}
where $\vec{a}$ and $\vec{b}$ are vectors that are determined by imposing the
continuity of the wave functions and their derivatives at $x=0$ and $x=L_x$.
Introducing the matrices $\mathsf{Q}=\mathsf{D}^{1/2}$,
$(\mathsf{K})_{n,n'}=k_n\delta_{n,n'}$,
$\mathsf{X}=\exp(i\mathsf{Q}L_x)$, $\mathsf{Z}=\exp(i\mathsf{K}L_x)$
and the vectors $\vec{\delta}_0=(1,0,\ldots,0,\ldots)$,
$\vec{r}=(r_0,r_1,\ldots,r_n,\ldots)$, $\vec{t}=(t_0,t_1,\ldots,t_n,\ldots)$,
we find:
\begin{eqnarray}
\vec{\delta}_0+\vec{r} & = & \mathsf{P} (\vec{a}+\vec{b}\,), \,\,\,\,
\mathsf{K} (\vec{\delta}_0-\vec{r})  =  \mathsf{P}  \mathsf{Q}  (\vec{a}-\vec{b}\,), \label{eq24b}\\
\mathsf{Z}  \vec{t} & = & \mathsf{P} (\mathsf{X} \vec{a}+\mathsf{X}^{-1} \vec{b}\,), \,\,\,\,
\mathsf{K} \mathsf{Z} \vec{t}  =  \mathsf{P} \mathsf{Q} (\mathsf{X}
\vec{a}-\mathsf{X}^{-1}  \vec{b}\,). \label{eq24d}
\end{eqnarray}
By combining this set of matrix equations one determines the
unknown vector 
\begin{equation}
\vec{r}  =
-[\mathsf{A}_+\mathsf{X}^{-1}\mathsf{B}_++\mathsf{A}_-\mathsf{X}\mathsf{B}_-]^{-1}[\mathsf{A}_-\mathsf{X}\mathsf{B}_++\mathsf{A}_+\mathsf{X}^{-1}\mathsf{B}_-]\vec{\delta}_0,
\label{appendixB2}
\end{equation}
and infer from it the other unknown vectors 
\begin{equation}
\fl \vec{a}  =
(\mathsf{B}_+\vec{\delta}_0+\mathsf{B}_-\vec{r}\,)/2, \,\,\,
 \vec{b} =(
\mathsf{B}_-\vec{\delta}_0+\mathsf{B}_+\vec{r}\,)/2, \,\,\, {\rm and}  \,\,\,
\vec{t}  =
\mathsf{Z}^{-1}\mathsf{P}(\mathsf{X} \vec{a} +  \mathsf{X}^{-1} \vec{b}\,),
\label{appendixB3}
\end{equation}
with $\mathsf{A}_\pm = \mathsf{K}\mathsf{P}\pm
\mathsf{P}\mathsf{Q}$ and $\mathsf{B}_\pm =\mathsf{P}^{-1} \pm
\mathsf{Q}^{-1} \mathsf{P}^{-1}\mathsf{K}$.

In practice, it is important to take into account a sufficient number of evanescent modes to ensure the convergence of the result. 
For the results of Fig.~\ref{fig4}, the number of propagating modes is $N_p=3$ (modes 0,2,4) but we have solved the equations detailed in this appendix 
using 25 modes since the convergence on the probabilities $R_i$ and $T_i$ was obtained with typically 15 modes for our parameters.

\section*{References}
\bibliography{BibArticle}
\end{document}